# Electron-Hole Symmetry Breaking in Charge Transport in Nitrogen-Doped Graphene

*Jiayu Li[†,§,#], Li Lin[‡,#], Dingran Rui[†], Qiucheng Li[‡], Jincan Zhang[‡], Ning Kang[†,\*], Yanfeng Zhang[‡,∥], Hailin Peng[‡], Zhongfan Liu[‡,\*], and H. Q. Xu[†,¤,\*]*

[†]Bejing Key Laboratory of Quantum Devices, Key Laboratory for the Physics and Chemistry of Nanodevices, and Department of Electronics, Peking University, Beijing 100871, P. R. China

[‡]Center for Nanochemistry, Beijing Science and Engineering Center for Nanocarbons, Beijing National Laboratory for Molecular Sciences, College of Chemistry and Molecular Engineering, Peking University, Beijing 100871, P. R. China

[§]Academy for Advanced Interdisciplinary Studies, Peking University, Beijing 100871, China

[∥]Department of Materials Science and Engineering, College of Engineering, Peking University, Beijing 100871, China

[¤]Division of Solid State Physics, Lund University, Box 118, S-22100 Lund, Sweden

[#]These authors contribute equally to this work.

\*Corresponding authors

Email: hqxu@pku.edu.cn (H. Q. Xu); Email: nkang@pku.edu.cn (N. Kang); Email: zfliu@pku.edu.cn (Z. Liu)

ABSTRACT: Graphitic nitrogen-doped graphene is an excellent platform to study scattering processes of massless Dirac fermions by charged impurities, in which high mobility can be preserved due to the absence of lattice defects through direct substitution of carbon atoms in the graphene lattice by nitrogen atoms. In this work, we report on electrical and magnetotransport measurements of high-quality graphitic nitrogen-doped graphene. We show that the substitutional nitrogen dopants in graphene introduce atomically sharp scatters for electrons but long-range Coulomb scatters for holes and, thus, graphitic nitrogen-doped graphene exhibits clear electron-hole asymmetry in transport properties. Dominant scattering processes of charge carriers in graphitic nitrogen-doped graphene are analyzed. It is shown that the electron-hole asymmetry originates from a distinct difference in intervalley scattering of electrons and holes. We have also carried out the magnetotransport measurements of graphitic nitrogen-doped graphene at different temperatures and the temperature dependences of intervalley scattering, intravalley scattering and phase coherent scattering rates are extracted and discussed. Our results provide an evidence for the electron-hole asymmetry in the intervalley scattering induced by substitutional nitrogen dopants in graphene and shine a light on versatile and potential applications of graphitic nitrogen-doped graphene in electronic and valleytronic devices.

Graphene, a honeycomb lattice of single layer carbon atoms with a Dirac-like energy dispersion, has attracted extensive interests in recent years owing to its extraordinary charge transport properties and potential applications in electronic devices.[1-5] However, charge impurities universally exist in graphene and influence its electronic properties.[6-10] A particular case is when approaching the Dirac point at which, because of the vanishing density of states, the transport properties of graphene are expected to be highly sensitive to scattering from charged impurities.[6-10] Previous studies of charge carrier scattering in graphene are mainly achieved with the scattering centers introduced by physical methods, such as ion irradiation,[7, 8] adsorption of adlayers,[9] low temperature deposition[10] and spin coating[11-13] of nanoparticles, and atomic hydrogen adsorption.[14] These methods could inevitably induce randomness and disorder in graphene and thereby largely degrade its transport properties. Compared with the physical methods, chemical doping has the advantage to achieve scattering centers in graphene with good replicability, lattice stability, and tunablity of doping concentration through the substitution of carbon atoms in the graphene lattice by dopant atoms. Among numerous dopants, nitrogen (N) atoms can be chemically incorporated into graphene forming covalent bonds with carbon (C) atoms. N-doped graphene has been employed for various applications including, *e.g.*, realizations of ultracapacitors,[15] lithium batteries,[16] and field-effect transistors.[17] Theoretical calculations and recent experiments have revealed that among the three predominant types of C-N bonding configurations (*i.e.*, graphitic N, pyridinic N, and pyrrolic N configurations) in graphene,[18-21] only graphitic N-doping (*i.e.*, substitutional N doping) could both bring n-type doping to graphene and preserve high mobility. (Here, it should be noted that other methods, such as atomic hydrogen adsorption, could also induce n-type doping in graphene.[14]) So far, several chemical approaches have been proposed to synthesize N-doped graphene.[22-28] However, the achieved N-doped graphene often shows a rather low carrier mobility compared with pristine graphene.[22-29] Here, it still remains as a challenging task to achieve N-doped graphene with controlled doping type and doping level and with a desired high carrier mobility.

Recently, we have succeeded in synthesizing large scale substitutional graphitic N-doped graphene with an ultra-high carrier mobility by means of chemical vapor deposition (CVD).[30] In this work, we report on an investigation of electron transport properties of our high-quality, graphitic N-doped graphene by electrical and magnetotransport measurements. Dominant scattering processes, such as intervalley scattering, intravalley scattering and phase coherent scattering, in graphitic N-doped graphene are studied and analyzed. We show that graphitic N-doped graphene exhibits a strong electron-hole asymmetry in the intervalley scattering and thus the transport characteristics. The physical origin of the asymmetry is deduced as that the substitutional N dopants in graphene act dominantly atomically sharp scatters for electrons but long-range Coulomb scatters for holes. We also demonstrate that charge carrier scattering in graphene can be effectively tuned by incorporation of different amount of substitutional N dopants in graphene and show that with increasing N atomic concentration, the strength of intervalley scattering is enhanced.

RESULTS AND DISCUSSION

The high-quality single-crystalline N-doped graphene used in this study was grown by CVD with a multiple-passivation growth strategy (see experimental section for details).[29] Different N doping levels were achieved by tuning growth temperature. In the CVD growth of N-doped graphene, due to the temperature-dependent competition in the formation of C-C and C-N bonds, a high growth temperature would lead to a larger probability to form more stable C-C bonds and thus a reduction in atomic N concentration in graphene lattices. To prove the capability to grow N-doped, high-quality graphene with different N atomic concentrations, the graphene films, formed at 1000 ºC, 950 ºC, 920 ºC and 900 ºC, were transferred onto $SiO_2$/Si substrates and were characterized by spectroscopic measurements.

X-ray photoelectron spectroscopy (XPS) measurements are informative with respect to the N atomic concentration and bonding type of dopants in the carbon lattice of graphene, in which the peak intensity (area) would reflect the dopant concentration and the position of the peaks the bonding environment. Figure 1a shows the XPS analysis for the N 1s peak of our sample gown at 900 °C (see Figure S1a for the XPS analyses of the N 1s peaks of the samples grown at three other temperatures). The XPS spectra show only one N 1s peak centred at 401.0 eV in each graphene sample, suggesting that there is only one N lattice configuration, *i.e.*, the substitutional doping type of N atoms (graphitic configuration), in the graphene samples.[28] Moreover, by taking the area ratio of the N 1$s$ and C 1$s$ peaks (the latter not shown here) after considering the atomic sensitivity factors, the atomic doping level can be estimated to be from 0.5% in the graphene sample grown at 1000 °C to 2.0 % in the graphene sample grown at 900 °C (Figure 1a and Figure S1a).

Scanning tunneling microscope (STM) measurements were performed and Figure 1b is an STM image of a typical as-synthesized N-doped graphene film. Here, the substitutional nature of N dopants (*i.e.*, the nature of atomically sharp point N impurities) in graphene is clearly displayed. As previously reported,[31] we could also observe long "tails" around each dopant, a signature of electron intervalley scattering induced by the N dopants. The inset in Figure 1b shows the fast Fourier transform (FFT) spectrum of the STM image. Clearly, the FFT spectrum shows two sets of points with each set being arranged as a hexagon. The points in the outer hexagon in the FFT spectrum arises from the atomic lattice of graphene and the points in the inner hexagon are derived from the points in the outer hexagon by intervalley scattering induced by N dopants, consistent with previous reports.[31-33] Thus, the intervalley scattering revealed by the STM analysis in our graphitic N-doped graphene is the results of scattering by the atomically sharp, substitutional N impurities.

In addition, Raman spectroscopy measurements of our as-synthesized graphene films were conducted. Figure 1c presents a representative Raman spectrum of an N-doped graphene film grown at 900 °C (see Figure S1b for the Raman spectra of the N-doped graphene films grown at three other temperatures). All our samples exhibit a prominent D band (Figure 1c and Figure S1b), whose intensity is reported to reflect the level of N-dopant concentration in the graphene sheets.[17, 34, 35] The intensities of the D band exhibit a clear reduction in the films grown at higher temperatures (Figure S1b). In addition, all the N-doped graphene films display a strong 2D band, indicating that the as-grown films are of high crystal quality (Figure 1c and Figure S1b).

For transport measurements, large-sized N-doped graphene films transferred onto $SiO_2$/Si substrates were selected and processed to form standard Hall bar structures. Raman spatial mapping for a typical fabricated Hall bar device is shown in the inset of Figure 1c. The observed uniform Raman D-band intensity over the entire Hall bar indicates a large-scale dopant homogeneity in the N-doped graphene sample. Figure 1d shows the longitudinal resistance measured as a function of gate voltage for an N-doped graphene sample with the N atomic concentration of 2% at both room temperature and 1.9 K. From the transfer characteristic curves, we can extract the carrier mobility in the graphene sample. The extracted values of the electron and the hole mobility at 1.9 K are ~8500 and ~8950 $cm^2 V^{-1} s^{-1}$, respectively. At room temperature, a value of ~8000 $cm^2 V^{-1} s^{-1}$ is extracted for the electron and the hole mobility. The well-defined quantum Hall plateaus (Figure S2a) and Shubnikov-de Hass oscillations (Figure S2b) are observed in the N-doped graphene sample, which further confirms the high mobility of our sample, compared to those previously reported for N-doped graphene.[26-28]

Figure 2a displays the conductivity measured as a function of back gate voltage $V_g$ for N-doped graphene samples with three different N atomic concentrations at 1.9 K and zero magnetic field. In a graphitic N-doped graphene sample, an N atom serves as a donor and becomes positively charged after sacrificing an electron. As expected, a graphitic N-doped graphene sample shows n-type transport characteristics, with the Dirac point located at negative gate voltages. As shown in Figure 2a, with an increase in N atomic concentration, the minimum conductivity point shifts toward a larger negative value in gate voltage. It is also seen that tuning N atomic concentration is accompanied by a change in the density of charged impurities. In order to quantitatively estimate the density of the charged impurities in our N-doped graphene samples, we use a Boltzmann kinetic theory,[20] considering charged impurity scattering in graphene, to fit our measured conductivity data. In the theory, the conductivity can be written as:[20]

$$\sigma(n) = \frac{20e^2}{h}\left|\frac{n}{n_{imp}}\right| + \sigma_{res} ,$$

where $h$ is the Plank constant, $e$ is the electron charge and $\sigma_{res}$ is the residual conductivity at the Dirac point. Figure 2b shows the extracted density of charged impurities $n_{imp}$ as a function of N atomic concentration. It is seen that $n_{imp}$ increases with increasing N-atomic concentration, as expected. Besides, the estimated effective N doping density from the N atomic concentration is consistent with the extracted charged impurity density, indicating that the positively charged N atoms are the main source of charged impurities in the graphitic N-doped graphene samples. We also estimated the mobility of the samples from the gate dependence of the conductivity at different N atomic concentrations. As one can see from Figure 2c, the mobility decreases as the N atomic concentration increases, which can be qualitatively understood as that more atomic scattering centers have been introduced to graphene at a higher N-atomic concentration.

We now consider the effect of N doping on the magnetotransport properties of graphene. Magnetotransport measurements have been widely used to probe carrier transport processes such as scattering mechanisms and quantify various characteristic length scales in graphene.[36-38] The chirality of graphene would give rise to the destructive interference of charge carriers in graphene at zero magnetic field. Thus, an increase in the conductivity, namely weak antilocalization (WAL), is expected to occur in monolayer graphene as long as the chiral symmetry is preserved during coherent backscattering.[36] However, in addition to inelastic phase-breaking scattering, elastic intervalley scattering, induced by atomically sharp defects or local deformation,[39] can break the chirality properties of the charge carriers, leading to suppression of WAL, and restore conventional weak localization (WL) in graphene.[37, 38]

Previous experimental findings and theoretical calculations have suggested that N-doping plays an important role in the intervalley scattering in graphene.[31, 39, 40] However, the effects of the N doping on transport properties have received little attention so far. Figure 3a shows the transfer curve of the sample with the 2% N atomic concentration measured at 1.9 K, in which the gate voltage at the Dirac point ($V_{Dirac}$ ~ -16V) has been shifted to zero for clarity. In Figure 3b, we plot the measured magnetoconductivity as a function of the magnetic field applied perpendicular to the graphene plane for selected gate voltages, on both the hole and electron transport sides, as indicated by color marks on the $R$-$V_g$ curve in Figure 3a. In Figure 3b, the curve measured at $\Delta V_g = -23$ V is placed on top and all the other measured curves are successively offset vertically by $-0.5 e^2/h$ for clarity.

A positive magnetoconductance with an obvious dip is clearly observed at all the measured curves, which can be ascribed to the effect of WL due to the presence of strong intervalley scattering in the N-doped graphene sample. As the gate voltage is tuned away from the Dirac point, the magnetoconductivity dip at zero magnetic field becomes sharper, which indicates an increase in phase coherence length with increasing carrier density on both hole and electron transport sides, in agreement with the previous experiments on graphene.[37, 38] The most striking feature seen in Figure 3b is that the magnetoconductivity dips at zero magnetic field on the hole transport side are sharper than their corresponding ones on the electron transport side. While in pristine graphene without N-doping, the measured magnetoconductivity dips at zero magnetic field show similar behaviors on the electron and hole transport sides (Figure S3). To clarify the physical origin of this asymmetry in the magnetoconductivity in the N-doped graphene samples, we analyze the measured results using the theory of quantum interference in monolayer graphene,[39] which shows that the magnetoconductivity is given by

$$\Delta\sigma(B) = \frac{e^2}{\pi h} \times \left[ F\left(\frac{B}{B_\varphi}\right) - F\left(\frac{B}{B_\varphi + 2B_i}\right) - 2F\left(\frac{B}{B_\varphi + B_i + B_*}\right) \right], \tag{1}$$

with

$$F(z) = \ln(z) + \Psi\left(\frac{1}{2} + \frac{1}{z}\right) \text{ and } B_{\varphi,i,*} = \frac{\hbar}{4De\tau_{\varphi,i,*}},$$

where $\Psi$ is the digamma function, $\tau_\varphi$ the phase coherence time, $\tau_i$ the intervallley scattering time, and $\tau_*$ the intravalley scattering time. The best fits of the measured data to the theory [Eq. (1)] are presented by the solid lines in Figure 3b. The scattering rates, $\tau_\varphi^{-1}$, $\tau_i^{-1}$ and $\tau_*^{-1}$, extracted from the fits are plotted in Figure 3c. It is seen that the intravalley scattering rate $\tau_*^{-1}$ shows small changes with the gate voltages. This is in agreement with the previous studies on graphene.[37]

A remarkable feature of the data presented in Figure 3c is that, as the gate voltage is swept through the Dirac point from the hole transport side to the electron transport side, the intervalley scattering rate $\tau_i$ rises sharply from ~1 ps$^{-1}$ to ~4 ps$^{-1}$, showing that the intervalley scattering is strongly asymmetric with respect to carrier type. In graphitic N-doped graphene, incorporation of N atoms into the graphene lattice is expected to give rise to atomically sharp point impurities, which could lead to an enhancement in intervalley scattering.[41] On the other hand, a substitutional N atom also serves as a donor and thus, after sacrificing an electron, acts as a positively charged scattering center for carrier transport in the graphitic N-doped graphene sample. Thus, on the hole transport side, due to Coulomb repulsive potential, the positively charged N impurities can deflect hole carriers,[42] leading to carrier scattering characterized dominantly by small-angle events. In contrast, on the electron transport side, electrons are attracted to positively charged N impurities due to Coulomb attraction. This Coulomb attraction favors to make electrons move closer to the impurities and then scattered off from the impurities dominantly by large-angle scattering events.[42] It is this kind of atomically sharp impurity scattering process that enhances the probability of intervalley scattering for electrons in our N-doped graphene samples. This asymmetry in the electron and hole intervalley scattering caused by the charged impurities has also been observed before in the transport and scanning tunneling microscope measurements of graphene decorated with metal atoms.[7, 41] It is worth to note that the measured transfer curves of our N-doped graphene samples also exhibit a noticeable electron-hole asymmetry. In Figure 2a, the conductivity in the hole transport regime shows a linear dependence on the gate voltage, which demonstrates that the ionized N impurities act dominantly as long-range Coulomb scatters for holes as discussed in previous works.[6, 43, 44] On the other hand, $\sigma$ ($V_g$) in the electron transport regime exhibits a weak sublinear dependence. Such a sublinear dependence has also been widely observed in previous works and has been attributed to the effect of short-range scatters in graphene, in consistence with our interpretation.

To explore the role of doping level in intervalley scattering in graphene, we have performed the magnetotransport measurements for a series of samples with different N atomic concentrations. Figure 3d displays the extracted intervalley scattering rate as a function of gate voltage for samples with N atomic concentrations of 1% (sample #85, Figures S4a and S4b) and 2% (sample #12). Both samples display electron-hole asymmetry in the intervalley scattering rate. In addition, the intervalley scattering rate in the sample with the N atomic concentration of 2% is approximately two times larger than that in the sample with the N atomic concentration of 1%, indicating that increasing N atomic concentration leads to an enhancement in intervalley scattering. To illustrate further the effect of doping level on the intervalley scattering rate, the measurements of the magnetoconductivity of the samples with four different N atomic concentrations (2.0%, 1.8%, 1.0%, and 0.5%) but the same carrier density ($n \sim 1\times10^{15}$ m$^{-2}$) are shown in Figure 3e. It is seen that, as the N atomic concentration increases, the magnetoconductivity dip at zero magnetic field becomes deeper. In Figure 3f, we plot the values of the intervalley scattering rate, extracted from fitting the measured magnetoconductivity data in the four samples to Eq. (1), as a function of the N atomic concentration. The intervalley scattering rate shows a monotonous increase with increasing N atomic concentration. Our results thus demonstrate that N doping is an efficient method to tune intervalley scattering in graphene.

Figure 4a displays the measured low-field magnetoconductivity for sample #12 (2% N atomic concentration) on the electron transport side at the gate voltage of $\Delta V_g$ =1V ($n \sim 1.8 \times 10^{15}$ m$^{-2}$) and different temperatures. At temperature $T$ = 1.9 K, we observe a positive magnetoconductivity, indicating a dominant WL behavior. With increasing temperature, the depth of the conductivity dip decreases gradually and the WL feature eventually disappears at temperatures of above 60 K. We have carried out the same analysis by fitting the magnetoconductivity data to Eq. (1). The results of the fits are presented by the solid lines in Figure 4a. Figure 4b shows the characteristic scattering lengths $l_\varphi$, $l_i$ and $l_*$, extracted from the fits, as a function of temperature, where $l_{\varphi,i,*} = \sqrt{D\tau_{\varphi,i,*}}$ and $D$ is the diffusion constant, $D = hv_F/4e^2\rho_{xx}\sqrt{\pi n}$, with $h$ being the Plank constant, $e$ the electric charge, $v_F$ the Fermi velocity, $\rho_{xx}$ the resistivity, and $n$ the carrier density extracted from the Hall measurements at low magnetic fields (Figure S5). Both the intervalley and intravalley scattering rates are found to be relatively temperature-independent, in agreement with previous reports.[37, 38] In contrast, the phase coherence length $l_\varphi$ increases with decreasing temperature at relatively high temperatures and tends to saturate at $T$ < 10 K. We also measured the temperature dependence of the magnetoconductivity at gate voltages $\Delta V_g$ = -7 V, -1 V, and 7 V (Figure S6). In Figure 4c, the extracted intervalley scattering rate is plotted as a function of temperature for gate voltages $\Delta V_g$ = -7 V, -1 V, 1 V, and 7 V. As expected, the intervalley scattering rate is roughly temperature-independent at all these gate voltages. But, again, it is seen that the intervalley scattering rate on the hole transport side is smaller than that on the electron transport side, exhibiting the clear electron-hole asymmetry as discussed above.

In Figure 4d, the dephasing rate obtained on both the electron and hole transport sides is shown as a function of temperature. As we can see, the dephasing rates of two types of charge carriers show different temperature dependences, suggesting that dominant dephasing scattering mechanisms of electrons and holes in our N-doped graphene samples are different. At low temperatures, the main source of phase-breaking in graphene is considered to be electron-electron interaction in two mechanisms.[38, 45, 46] One is the Nyquist scattering ($\tau_N^{-1}$) representing the small energy-transferred interactions of charge carriers with electromagnetic field fluctuations generated by the noisy movement of neighboring charge carriers.[47, 48] The other is the direct Coulomb interaction ($\tau_{ee}^{-1}$) among the charge carriers.[38, 45, 46] By considering both mechanisms,[37, 48] the dephasing rate can be written as

$$\frac{1}{\tau_\varphi} = \frac{1}{\tau_N} + \frac{1}{\tau_{ee}} + \text{const} = ak_BT\frac{\ln(g)}{\hbar g} + b\frac{\sqrt{\pi}}{2v_F}\left(\frac{k_BT}{\hbar}\right)^2\frac{\ln(g)}{\sqrt{n}} + \text{const},$$

where $g(n)=\sigma h/e^2$. The linear temperature dependent term corresponds to the inelastic scattering with small momentum transfer, while the parabolic temperature dependent term is due to large momentum transfer by the direct Coulomb interaction between charge carriers.[47, 48] Figure 4d (see also Figure S7) clearly displays a linear temperature dependence of the dephasing rate on the hole transport side, whereas a parabolic temperature dependence is observed on the electron transport side, suggesting that the direct Coulomb interaction between electrons is dominated on the electron transport side. In a simple picture, the accumulation of electrons around the positively charged N atoms could give rise to an enhancement in direct electron-electron interaction. Therefore, the evidences observed for different phase-breaking mechanisms with respect to carrier type further support the charged impurity-induced electron-hole asymmetry scenario in our N-doped graphene samples.

We now discuss about the transport behavior of N-doped graphene at high magnetic fields. Figure 5a shows the measured longitudinal resistance $R_{xx}$ of sample #02 (1.8% N atomic concentration) as a function of gate voltage at a perpendicularly applied magnetic field $B = 8$ T and different temperatures between 1.9 K and 250 K. It is seen that the $R_{xx}$ minima at filling factors $v = -2$ and $v = -6$ on the hole transport side are close to zero at $T = 1.9$ K, manifesting that the $v = -2$ and $v = -6$ quantum Hall (QH) states are well developed at the magnetic field. In contrast, the $R_{xx}$ minima on the electron transport side show distinct deviations from zero at the magnetic field. Furthermore, a shrinking of the Hall plateau in $R_{xy}$ at $v = +2$ on the electron transport side can also be seen in the inset of Figure 5a. The measurements clearly reveal that the QH states on the hole transport side are much more robust than the corresponding QH states on the electron transport side, suggesting the existence of different broadenings in the hole and electron Landau levels.

To quantitatively estimate the broadenings of the Landau levels, we perform activation gap measurements of the QH states at the minima of $R_{xx}$ where the Fermi level is in the middle of a gap between Landau levels and the thermal excitation of charge carries at elevated temperatures across the gap gives rise to finite bulk transport. As shown in Figure 5a, all the minima of $R_{xx}$ measured at different temperatures display a clearly activated transport behavior. Figure 5b shows the measured resistance minima at filling factors $v = -2$ and $v = +2$ as a function of temperature at $B = 8$ T. The temperature dependence of $R_{xx}$ can be described by[49-51]

$$R_{xx} \propto \exp(-\Delta/k_B T),$$

where $\Delta$ is the activation gap. From the slopes in the Arrhenius plots shown in Figure 5b, we can deduce the activation gap values of $\Delta/k_B = 636.8$ K at $v = -2$ and $\Delta/k_B = 374.9$ K at $v = +2$. This difference in activation gap indicates that the QH states on the hole transport side are more robustly defined than the QH states on the electron transport side (see also Figure S8 for the activation gap measurements of sample #12 with the 2% atomic N concentration) and the Landau level broadenings are larger on the electron transport side than on the hole transport side. This carrier type dependent difference in Landau level broadening provides another evidence for the electron-hole asymmetry in intervalley scattering in graphitic N-doped graphene. As we already showed in the low-field magnetotransport measurements, the major difference in carrier scattering between the electron and the hole transport is intervalley scattering in graphitic N-doped graphene. It is this difference that leads to the difference in large angle scattering and thus the dominant difference in the Landau level broadening of electrons and holes in our graphitic N-doped graphene samples.

CONCLUSIONS

In this work, electron-hole asymmetry in charge transport properties of N-doped graphene have been studied. The high-quality N-doped graphene films with different N atomic concentrations have been synthesized and standard Hall bar structures fabricated from the synthesized films have been electrically measured at different temperatures and magnetic fields. Scattering properties of the charge carriers in the N-doped graphene films are extracted from magnetotransport measurements at low magnetic fields. It is shown that the phase coherent scattering time and intravalley scattering time are approximately insensitive to the type of transport carriers in the N-doped graphene films. However, the intervalley scattering time is strongly carrier type dependent—the intervalley scattering is much stronger for electrons than for holes in the films. In addition, the intervalley scattering rate is found to increase with increasing N atomic concentration in the graphitic N-doped films, but it is approximately temperature independent. The observed electron-hole asymmetry in the intervalley scattering can be attributed to scattering by charged N dopants in the graphene films. In graphitic N-doped graphene films, N atoms act as positively charged impurities. Thus, electrons see them as attractive scattering centers and can be scattered off from them with a large probability by large-angle events and therefore by intervalley scattering processes. In contrast, holes see the positively charged N impurities as repulsive scattering centers and can dominantly be scattered off from them by small angle processes. We have also analysed phase coherent scattering of the carriers in the films at different temperatures and found that the contribution from direct electron-electron interaction to phase coherent breaking is more significant on the electron transport side than on the hole transport side. Furthermore, the transport measurements have been carried out for the films in the quantum Hall regime and it is extracted that the Landau level broadening is larger on the electron transport side than on the hole transport side. All these observations can be attributed to the presence of positively charged N impurities in the N-doped graphene films.

EXPERIMENTAL SECTION

**Graphene growth and transfer.** N-doped graphene was grown on commercially available Cu foils in a low pressure CVD system.[30] As-formed individual N-doped graphene domains or continuous films were transferred onto $SiO_2$/Si substrates with PMMA-assisted dry method for Raman spectroscopy characterization and electrical device fabrication.[52] Graphene was formed on both sides of a Cu foil, and graphene on one side of the Cu foil to be used for the characterization and device fabrication was spin-coated with PMMA. After baking at 150 °C for 5 min, the other side of the Cu foil was exposed to $O_2$ plasma for 3 min to remove graphene grown on it. Subsequently, the 1 M $Na_2S_2O_8$ solution was chosen to etch away the Cu foil. Then, the achieved floating PMMA/graphene membrane on the solution surface was washed with deionized water three times. To minimize the transfer-related contamination and defects, the sample was heated at 150 °C for 1 h. After drying, the PMMA was dissolved by acetone yielding graphene domains or continuous films on the substrate. Note that, in the dry transfer method, after rinsed by deionized water, the PMMA/graphene membranes were subsequently washed by isopropanol and then dried in air for 12 h before it was placed onto the target substrates.

**Device fabrication and transport measurements.** For device fabrication, as-formed N-doped graphene was transferred onto a highly doped Si substrate with 300-nm-thick $SiO_2$ on top. After a heating cleaning process at 150 °C, the samples were investigated with an atomic force microscope (AFM) to check the flatness of the surface. After that, the graphene samples were etched into standard Hall bars (with a width of 2 µm and a length of 6 µm) by electron beam lithography (EBL) and reactive ion etching. Once the samples were patterned, AFM imaging was performed again to ensure that the Hall bar regions were free from winkles or residues. Finally, Ti/Au (10 nm/90 nm) contact electrodes were fabricated by EBL and electron-beam evaporation.

Electrical and magnetotransport measurements were performed in a physical property measurement system (PPMS, Quantum Design DynaCool) equipped with a superconducting magnet capable of applying a magnetic field up to 9 T. The resistance of the samples was measured using a current-biased low-frequency (17.77 Hz) lock-in technique with an ac excitation current in the range of 0.01 to 1 μA. The longitudinal and the Hall resistance were simultaneously recorded during the whole magnetotransport measurements. The magnetic field was always applied perpendicular to the sample plane.

ASSOCIATED CONTENT

**Supporting Information.**

The Supporting Information is is available free of charge on the ACS Publication website http://pubs.acs.org. Supplementary Text, Supplementary Figures S1−S8, and additional References.

AUTHOR INFORMATION

**Corresponding Author**

*Email: hqxu@pku.edu.cn (H. Q. Xu); *Email: nkang@pku.edu.cn (N. Kang); *Email: zfliu@pku.edu.cn (Z. Liu)**Author Contributions**

[#] J.Y.Li and L.Lin contributed equally to this work.

**Notes**

The authors declare no competing financial interest.

ACKNOWLEDGMENTS

We thank Z. Jia, B. Yan and S. Li at Peking University for helpful discussions. This work was financially supported by the Ministry of Science and Technology of China (MOST) through the National Key Research and Development Program of China (No. 2016YFA0300601) and the National Basic Research Program of China (Nos. 2012CB932703 and 2012CB932700), and by the National Natural Science Foundation of China (Nos. 91221202, 91421303, and 11374019).

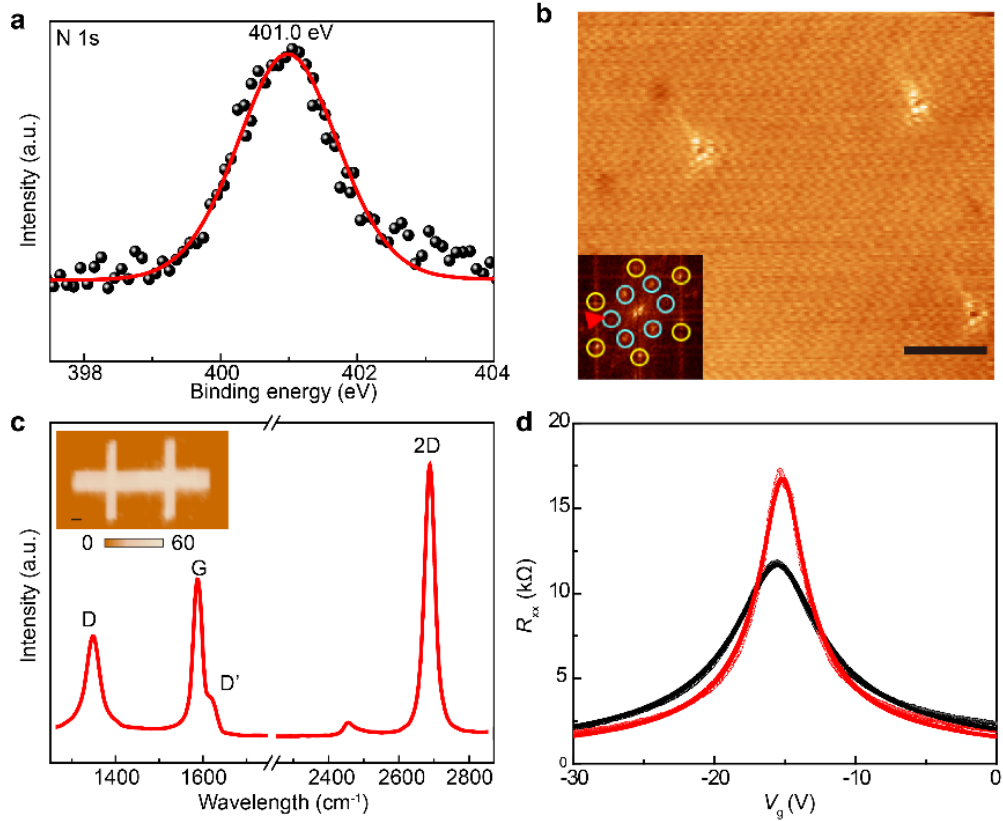

**Figure 1.** Characterizations of as-formed N-doped graphene films. (a) N 1s core-level XPS spectra of an as-formed graphitic N-doped (2% N atomic concentration) graphene film. The single N 1s peak, centered at 401.0 eV, justifies that substitutional doping is the doping configuration in the graphene lattice. (b) STM image of an as-formed graphitic N-doped graphene sample ($V_{bias}$ = -10 mV and $I_{set}$ = 100 pA). The bright spot patterns show single graphitic type of N dopant atoms in graphene. The inset is the FFT of the topography which shows diffraction peaks of the graphene lattice (outer hexagonal points, indicated by the yellow circles) and peaks induced by intervalley scattering (inner hexagonal points, indicated by blue circles). Scale bar is 2 nm. (c) Raman spectra of a graphitic N-doped (2% N atomic concentration) graphene film. The inset shows the integrated intensity Raman mapping of the D band over a Hall bar sample made from the graphitic N-doped graphene. Scale bar is 1 μm. (d) Resistivity of the graphitic N-doped graphene sample measured as a function of the back-gate voltage at both 300 K (black) and 1.9 K (red).

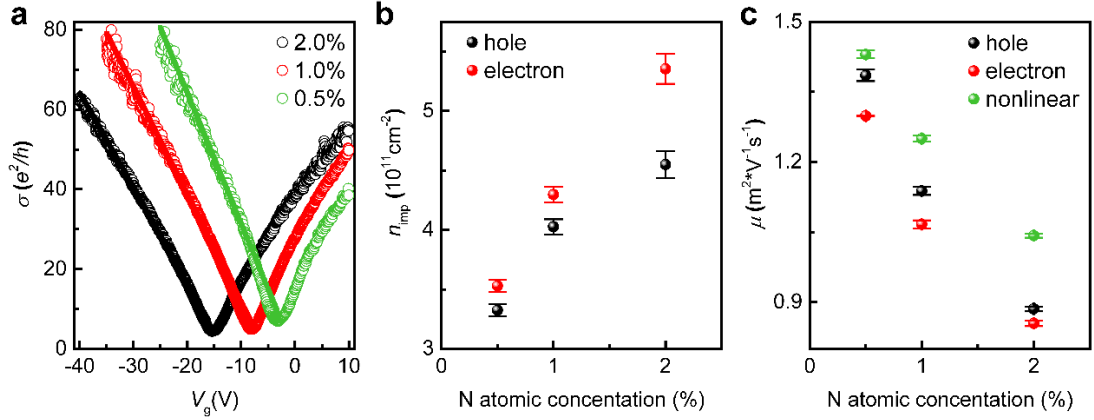

**Figure 2.** Transport properties of as-formed N-doped graphene films. (a) Conductivity ($\sigma$) *versus* gate voltage ($V_g$) taken at 1.9 K for the films with 0.5% (black symbols), 1.0% (red symbols), and 2.0% (green symbols) N atomic concentrations. The Dirac point is shifted towards a more negative value when the atomic N concentration is increased. (b) Charge impurity density *versus* atomic concentration extracted from the transport measurements on both the electron transport side and the hole transport side. The extracted charge impurity density increases with increasing atomic concentration. (c) Values of the carrier mobility extracted by linear fitting to the measured conductivity data on the hole transport side and on the electron transport side and by nonlinear fitting to the data over the entire region of the measurements. The extracted mobility decreases with increasing atomic N concentration.

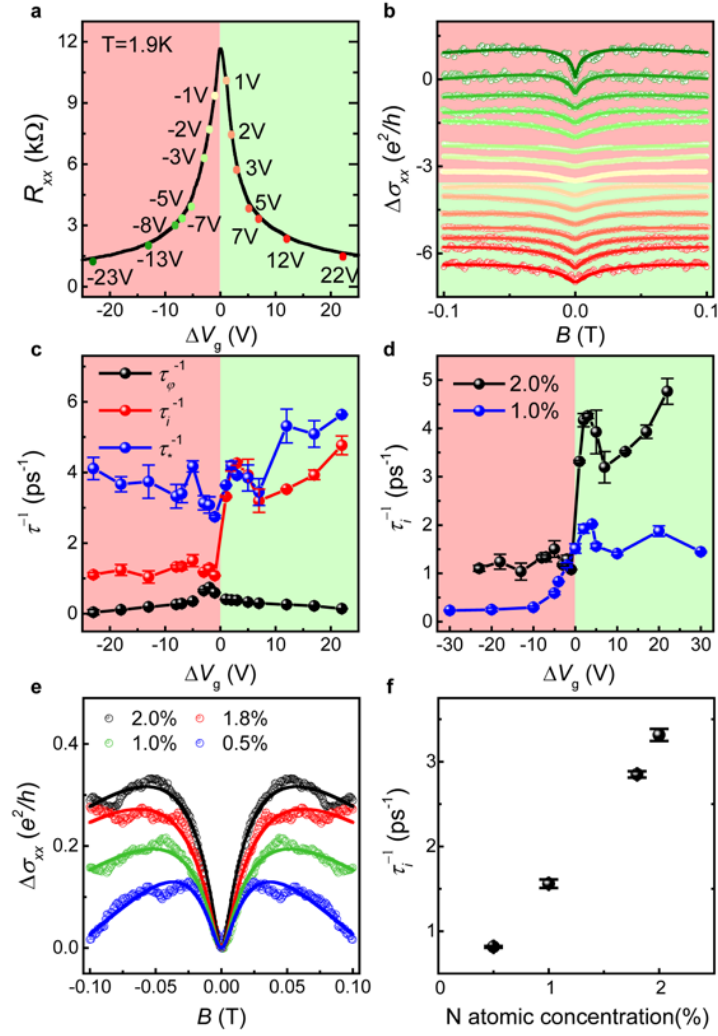

**Figure 3.** Low-field magnetotransport measurements of N-doped graphene samples at 1.9 K. (a) Shifted transfer curve of a N-doped graphene sample (sample #12) with a 2% N atomic concentration. In the originally measured transfer curve, the Dirac point is at $V_{Dirac}$ = -16V. The regions shaded by different colors represent the regions of the transport with different types of charge carriers. The color-highlighted dots on the shifted transfer curve represent different gate voltages at which the measured magnetoconductivity is to be plotted. (b) Magnetoconductivity measured for sample #12 at selected back gate voltages as indicated by the colored dots in the shifted transfer curve shown in (a). The curve

measured at $\Delta V_g$ = -23 V is placed on top and all the other measured curves are successively offset vertically by $-0.5e^2/h$ for clarity. The solid lines show the best fits to the experimental data. (c) Scattering rates extracted for sample #12 at different back gate voltages. (d) Intervalley scattering rates for sample #12 and sample #85 (with 1.0% N atomic concentration and the Dirac point at $V_{Dirac}$ = -4 V) at different gate voltages. (e) Magnetoconductivity measured for samples #12, #02, #85, and #47 (with N atomic concentrations of 2.0%, 1.8%, 1.0%, and 0.5%, respectively) on the electron transport side with the same carrier density of $n \sim 1\times10^{15}$ m$^{-2}$. (f) Intervalley scattering rate extracted as a function of N atomic concentration from the measured magnetoconductivity curves shown in (e).

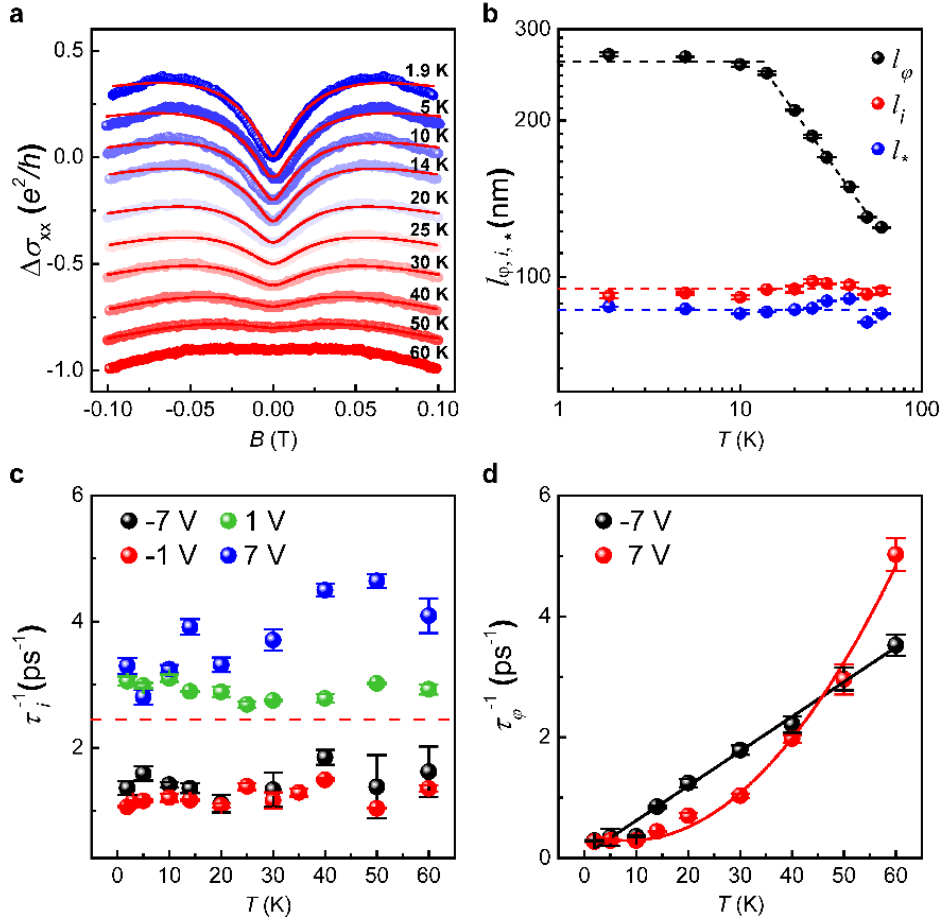

**Figure 4.** Low-field magnetotransport measurements at different temperatures. (a) Magnetoconductivity measured for sample #12 at $\Delta V_g = 1$ V and temperatures from 1.9 K to 60 K. The data measured at 1.9 K are placed on top and all other measured data are successively offset vertically by $-0.5 e^2/h$ for clarity. The solid lines show the best fits to the experimental data. (b) Characteristic scattering lengths, $l_\varphi$, $l_i$ and $l_*$, extracted from the measured magnetoconductivity as a function of temperature. (c) Scattering rates extracted for the sample at four different gate voltages and different temperatures. (d) Phase coherent breaking rates extracted for the sample at $\Delta V_g = 7$ V (the electron transport side) and $\Delta V_g = -7$ V (the hole transport side) at different temperatures. A linear (parabolic)

temperature dependence of the phase coherent breaking rate is seen on the hole (electron) transport side.

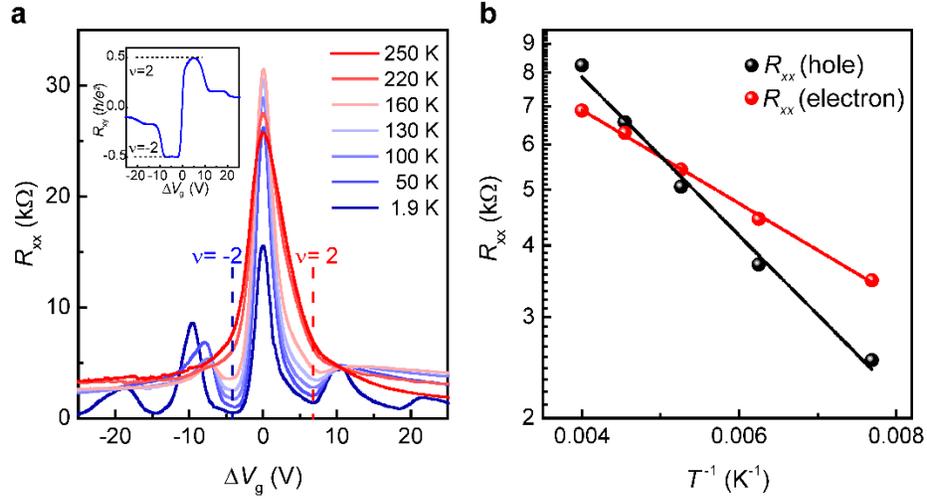

**Figure 5.** Activation gap measurements at a high magnetic field. (a) Transfer curves measured for sample #2 (with 1.8% N atomic concentration) at a magnetic field of 8 T and different temperatures (only seven out of thirty lines are shown here). (b) Arrhenius plots of $R_{xx}$ in an elevated temperature range for the sample at filling factors $v = -2$ and $v = 2$. The activation gaps ($\Delta$) extracted from the Arrhenius plots are $\Delta/k_B = 636.8$ K at $v = -2$ and 374.9 K at $v = 2$.

# Supporting Information for Electron-Hole Symmetry Breaking in Charge Transport in Nitrogen-Doped Graphene

*Jiayu Li[†,§,#], Li Lin[‡,#], Dingran Rui[†], Qiucheng Li[‡], Jincan Zhang[‡], Ning Kang[†,*], Yanfeng Zhang[‡,∥], Hailin Peng[‡], Zhongfan Liu[‡,*], and H. Q. Xu[†,¤,*]*

[†]Beijing Key Laboratory of Quantum Devices, Key Laboratory for the Physics and Chemistry of Nanodevices, and Department of Electronics, Peking University, Beijing 100871, P. R. China

[‡]Center for Nanochemistry, Beijing Science and Engineering Center for Nanocarbons, Beijing National Laboratory for Molecular Sciences, College of Chemistry and Molecular Engineering, Peking University, Beijing 100871, P. R. China

[§]Academy for Advanced Interdisciplinary Studies, Peking University, Beijing 100871, China

[∥]Department of Materials Science and Engineering, College of Engineering, Peking University, Beijing 100871, China

[¤]Division of Solid State Physics, Lund University, Box 118, S-22100 Lund, Sweden

[#]These authors contribute equally to this work.

*Corresponding authors
Email: hqxu@pku.edu.cn (H. Q. Xu); Email: nkang@pku.edu.cn (N. Kang); Email: zfliu@pku.edu.cn (Z. Liu)

**Estimation of the FET mobility and carrier density**

Nonlinear fitting method for the extraction of the FET mobility μ.

The total carrier density (electrons or holes) in graphene $n_{tot}$ can be approximated by[1]

$$n_{tot} = \sqrt{n^2(V_g) + n_0^2} \ ,$$

where $n_0$ represents the density of carriers at the Dirac point, often referred to as the residual carrier density. By ignoring the quantum capacitance, the gate induced carrier density can be calculated by

$$n(V_g) = \frac{C_{ox}*(V_g - V_{Dirac})}{e} \ ,$$

where $C_{ox}$ is the oxide capacitance, $V_{Dirac}$ is the back gate voltage at the Dirac point, $e$ is the electron charge. In the Hall bar geometry, the longitudinal resistance $R_{xx}$ at zero magnetic field can be written as

$$R_{xx} = \frac{L}{W} \frac{1}{e\,\mu\,\sqrt{n^2(V_g)+n_0^2}} \ ,$$

where $L$ is the distance between two voltage probes along the current direction, $W$ is the width of the Hall bar, and $\mu$ represents the field-effect transistor (FET) mobility.

Estimation of the carrier density by Hall measurements.

As shown in Supplementary Fig. S5, the Hall resistance $R_{xy}$ is a linear function of magnetic field $B$ at low magnetic fields at a given back gate voltage $V_g$ far from the Dirac point. The Hall coefficient $R_H$ can be obtained from the slope of the measured Hall resistance curve and the carrier density can be calculated as $n = 1/eR_H$.

**Temperature dependence of the phase coherent length**

We have shown that the phase coherent scattering rate exhibits a linear dependence on temperature on the hole transport side and a parabolic dependence on temperature on the electron transport side. Here, we note that, as shown in Supplementary Figs. S6(b), S6(d), and S6(f), as well as Fig. 4(b) in the main article, the phase coherent length $l_\varphi$ increases with decreasing temperature but shows a saturation behavior at temperatures below 10 K.

**Reference**

(1) Kim, S.; Nah, J.; Jo, I.; Shahrjerdi, D.; Colombo, L.; Yao, Z.; Tutuc, E.; Banerjee, S. K. Realization of a High Mobility Dual-Gated Graphene Field-Effect Transistor with $Al_2O_3$ Dielectric. *Appl. Phy. Lett.* **2009**, 94, 062107.

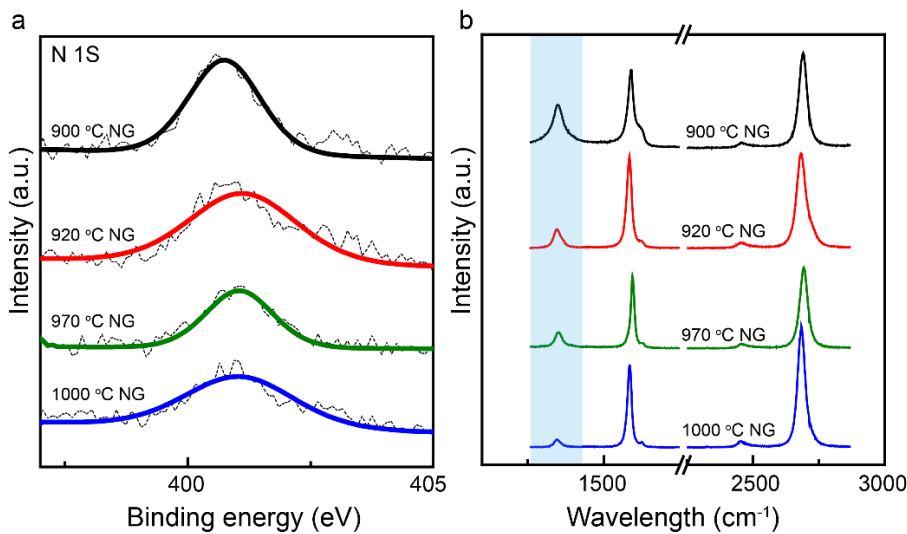

**Figure S1. Characterization of the doping concentrations in graphitic N-doped graphene films grown at different temperatures.** (**a**) N 1s XPS spectra of the graphitic N-doped graphene films grown at 900 ºC, 920 ºC, 950 ºC, and 1000 ºC. The N 1s peak is located at ~401 eV, indicating the dominance of the substitutional N atomic doping. Note that, by taking the peak area ratio of the N 1s at ~401 eV to the C 1s at ~285 eV after considering the atomic sensitivity factors, the percentage of N atoms in the graphene lattice is estimated to be 2.0%, 1.8%, 1.0%, 0.5% for the films grown at 900 ºC, 920 ºC, 950 ºC, and 1000 ºC, respectively. (**b**) Raman spectra of the films grown at 900 ºC, 920 ºC, 950 ºC, and 1000 ºC. Note that the D band regions in the spectra are shaded by light blue color in the figure and the intensity of the D bands reflects the concentration of dopants in the graphene lattice. Clearly, both the intensities of the D bands in the Raman spectra and the N 1s peaks in the XPS spectra decrease with increasing growth temperature, indicative of a clear reduction of the doping concentration in the N-doped films grown at high temperatures.

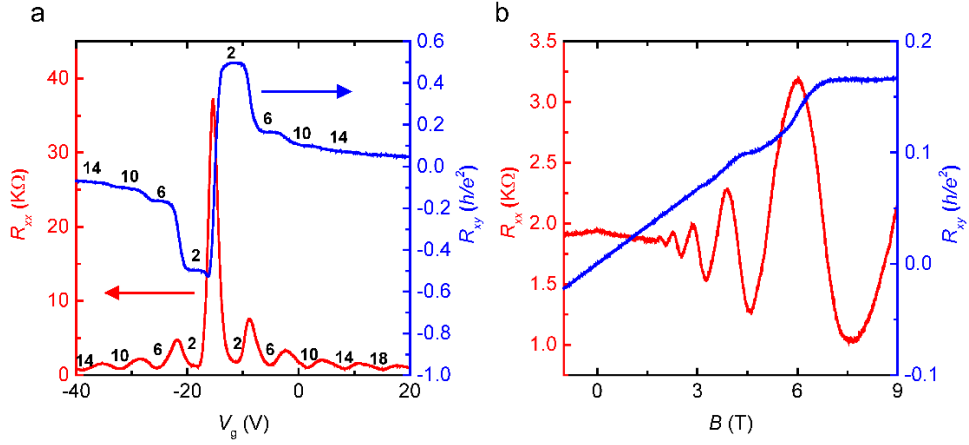

**Figure S2. Quantum Hall effect and Shubnikov−de Haas oscillations measured for a graphitic N-doped graphene sample (sample #12 with a 2% N atomic concentration).** (**a**) Longitudinal resistance ($R_{xx}$, red) and Hall resistance ($R_{xy}$, blue) measured for the graphene sample as a function of the back gate voltage at 1.9 K and a magnetic field of 5 T. The numbers in the figure indicate filling factors. (**b**) $R_{xx}$ and $R_{xy}$ measured for the sample as a function of magnetic field at a fixed gate voltage (0 V). Shubnikov−de Haas oscillations are observable in $R_{xx}$ at magnetic fields as low as 2 T, whereas $R_{xy}$ begin to reveal plateau-like structures at a magnetic field of ~3 T. The observation of well-developed quantum Hall plateaus and Shubnikov−de Haas oscillations at relatively low magnetic fields indicate high quality of our graphene samples.

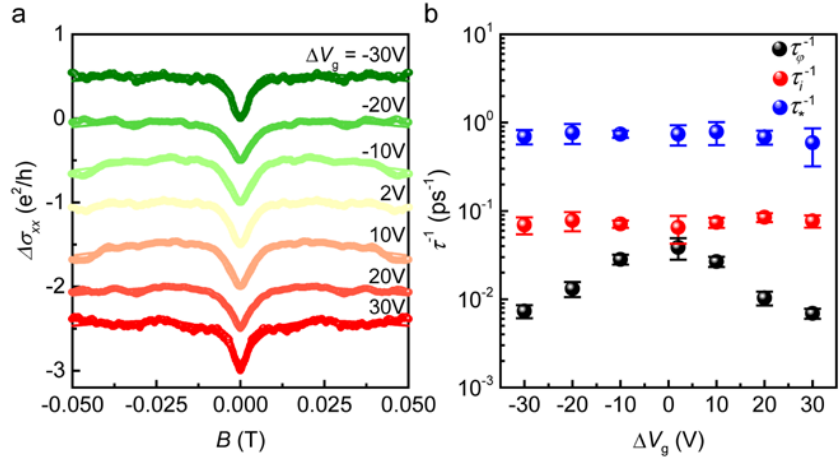

**Figure S3. Magnetoconductivity measurements of pristine graphene with zero nitrogen doping.** (**a**) Magnetoconductivity measured for a pristine graphene sample at different back gate voltages $\Delta V_g$ (colored solid dots). The data measured at $\Delta V_g$ = -30 V are plotted on top and all other measured data at different values of $\Delta V_g$ are successively offset vertically by $-0.5 e^2/h$ for clarity. The solid lines are the best fits to the experimental data. (**b**) Characteristic scattering rates of charge carriers extracted for the sample as a function of $\Delta V_g$, showing no clear electron-hole asymmetry in intervalley scattering rate.

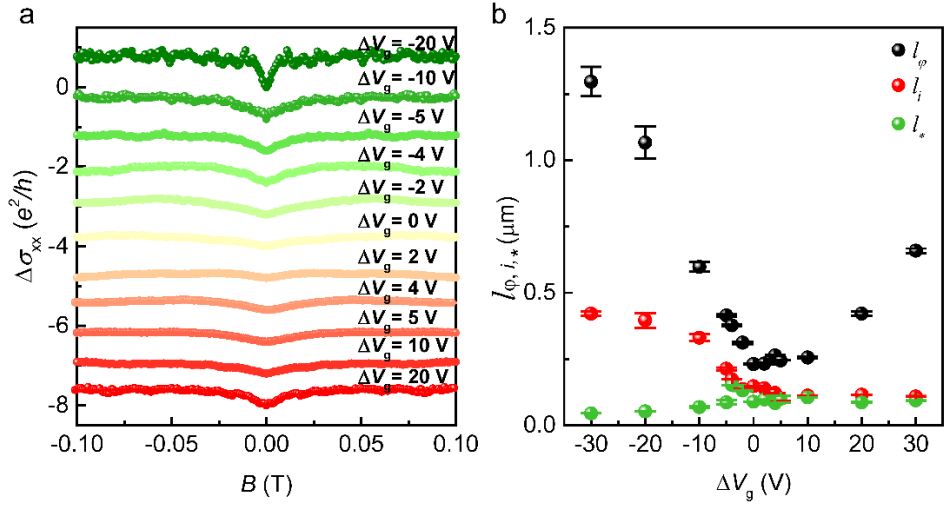

**Figure S4. Magnetoconductivity measurements of sample #85 (with a 1% N atomic concentration and the Dirac point at $V_{Dirac}$ = -10 V).** (**a**) Magnetoconductivity measured for the sample at different back gate voltages $\Delta V_g$ (colored solid dots). The data measured at $\Delta V_g$ = -20 V are plotted on top and all other measured data at different values of $\Delta V_g$ are successively offset vertically by $-0.8e^2/h$ for clarity. The solid lines are the best fits to the experimental data. (**b**) Characteristic scattering lengths of charge carriers extracted for the sample as a function of $\Delta V_g$, showing a clear electron-hole asymmetry in intervalley scattering length.

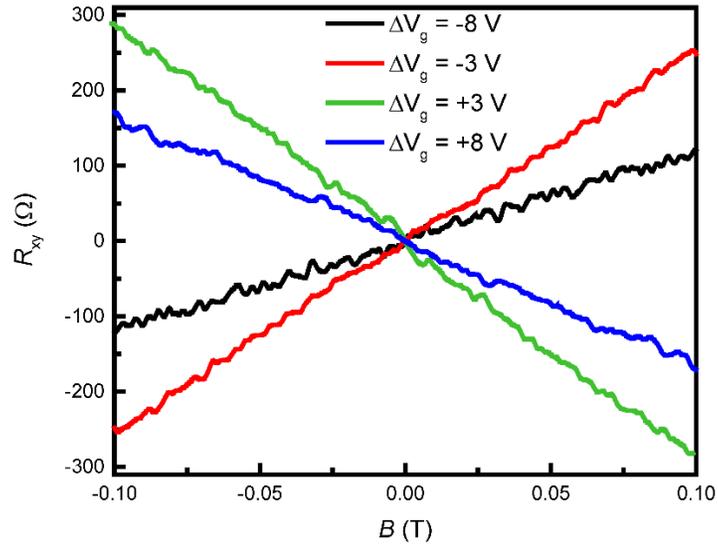

**Figure S5. Hall resistance measurements of sample #12 (with a 2% N atomic concentration) for extraction of the carrier density.** Hall resistance measured for the sample as a function of magnetic field at $T = 1.9$ K and four representative back gate voltages of $\Delta V_g$ = -8 V, -3 V, +3V and +8 V (measured from the Dirac point). All the measured traces can be fitted by straight lines and the carrier densities can be extracted from the slopes of the straight lines.

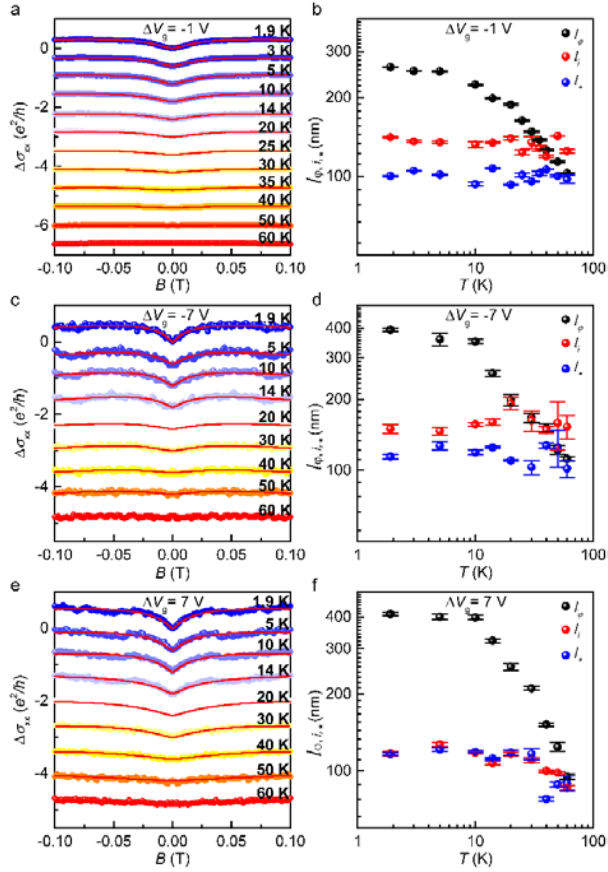

**Figure S6. Temperature dependent measurements of the magnetoconductivity of sample #12 (with a 2% N atomic concentration) at different back gate voltages.** (**a**) Magnetoconductivity of the sample measured at back gate voltage $\Delta V_g$ = -1 V and different temperatures (solid dots). The data measured at -1.9 K are plotted on top and all other data measured at different temperatures are successively offset vertically by $-0.6e^2/h$ for clarity. The solid lines show the best fits to the experimental data. (**b**) Characteristic scattering lengths extracted for the sample from the measurements shown in (a). (**c**) The same as (a) but for measurements at $\Delta V_g$ = -7 V. (**d**) The same as (b) but extracted from the measurements shown in (c). (**e**) The same as (a) but for measurements at $\Delta V_g$ = 1 V. (**f**) The same as (b) but extracted from the measurements shown in (e).

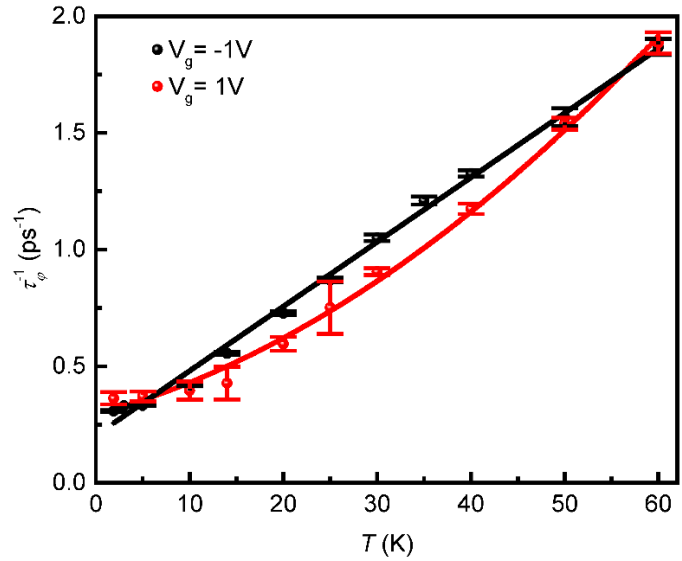

**Figure S7. Temperature dependence of the phase coherent breaking rate extracted for sample #12 (with a 2% N atomic concentration) at $\Delta V_g$= -1 V and 1 V.** The black solid dots are the experimental data extracted for the sample at $\Delta V_g$= -1 V from the measurements shown in Supplementary Fig. 6S(a) and the red dots are the experimental data extracted for the sample at $\Delta V_g$= 1V form the measurements shown in Fig. 4(a) in the main article. Here, as seen in the main article, a linear (parabolic) temperature dependence of the phase coherent breaking rate is seen on the hole (electron) transport side.

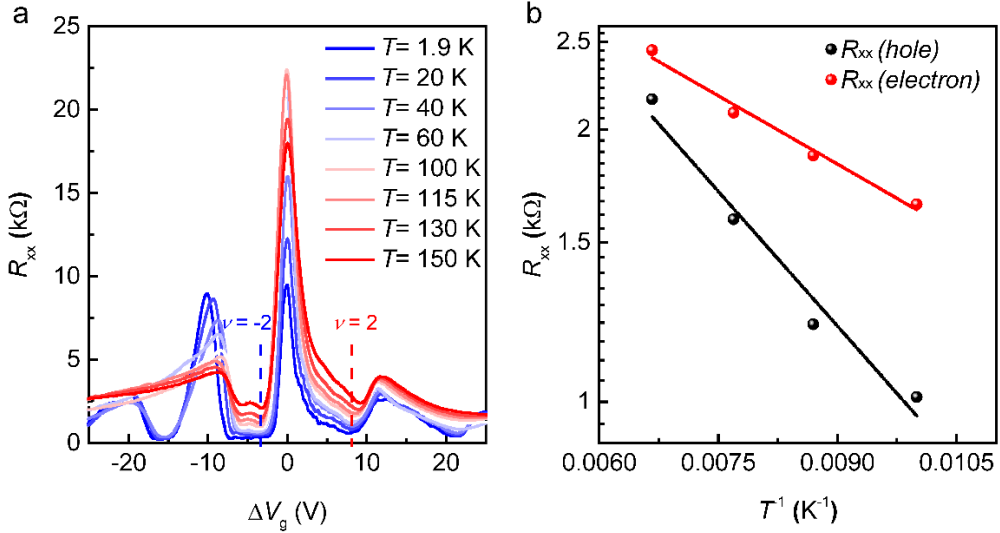

**Figure S8. Activation gap measurements for sample #12 (with a 2% N atomic concentration).** (**a**) Transfer curves measured for the sample at a magnetic field of 9 T and different temperatures (only eight out of thirty lines are shown here). (**b**) Arrhenius plots of $R_{xx}$ in an elevated temperature range for the sample at filling factors $v = -2$ and $v = 2$ as indicated in (a). The activation gaps ($\Delta$) extracted from the Arrhenius plots are $\Delta/k_B = 565.7$ K at $v = -2$ and 302.3 K at $v = 2$, showing the same electron-hole asymmetry in activation gap of the quantum Hall states as in the main article.